\documentclass[aps,prb,10pt,twocolumn,showpacs,superscriptaddress]{revtex4-1}
\usepackage{amsmath, amssymb, amsfonts}
\usepackage{graphicx}
\usepackage{bm}
\usepackage{dsfont}
\usepackage[squaren]{SIunits}
\usepackage{xcolor}
\usepackage[colorlinks=true,allcolors=blue]{hyperref}

\begin{document}

\title{Reservoir Computing with Random Skyrmion Textures}

\author{D.~Pinna}
\affiliation{Institute of Physics, Johannes Gutenberg-Universit{\"a}t, 55128 Mainz, Germany}

\author{G.~Bourianoff}
\affiliation{Intel Corp., retired}

\author{K.~Everschor-Sitte}
\affiliation{Institute of Physics, Johannes Gutenberg-Universit{\"a}t, 55128 Mainz, Germany}

\date{\today}

\begin{abstract}
The Reservoir Computing (RC) paradigm posits that sufficiently complex physical systems can be used to massively simplify pattern recognition tasks and nonlinear signal prediction. This work demonstrates how random topological magnetic textures present sufficiently complex resistance responses for the implementation of RC as applied to A/C current pulses. In doing so, we stress how the applicability of this paradigm hinges on very general dynamical properties which are satisfied by a large class of physical systems where complexity can be put to computational use. By harnessing the complex resistance response exhibited by random magnetic skyrmion textures and using it to demonstrate pattern recognition, we explain how spintronically accessible magnetic systems offer an advantage in the search for an ideal reservoir computer. The low-power properties of compact skyrmion fabrics, coupled with their CMOS integrability operating on similar length and timescales, open the door for their RC employment geared towards industrial application. 
\end{abstract}

\maketitle

%%%%% Article Body %%%%%%%%%%%%
\section{Introduction}
\textbf{Motivation --} It is widely acknowledged that classical Dennard scaling~\cite{Dennard1999} of complementary metal-oxide semiconductor (CMOS) logic, making devices simultaneously smaller, faster and operating at lower voltage, ceased at the turn of the millennium when transistor channel lengths reached the order of $100\,n\mathrm{m}$. Subsequently, dimensional scaling of semiconductors continued in step with Moore's law (with a two-year cadence), thus making devices smaller and cheaper. During that same period, a number of enhancements to CMOS were made to further improve transistor performance such as strained channels~\cite{ota2002novel}, high K/metal gates~\cite{auth200845nm}, steep sub-threshold devices~\cite{khatami2009steep}, trigate geometries~\cite{doyle2003high}, and high mobility channel materials~\cite{schwierz2010graphene,skotnicki2010can,del2011nanometre}. However, as we approach the theoretical limit of $3\,n\mathrm{m}$ transistor channel lengths~\cite{Zhirnov2003} manufacturing challenges become exponentially more difficult and more expensive to overcome. The increasing marginal cost of progressively smaller performance gains is expected to make further silicon scaling uneconomical. The latter is expected to occur within the next two fabrication generations with the advent of $\sim 5\,n\mathrm{m}$ transistor sizes.  Simultaneously, a seismic shift is occurring in the computational workloads from data-based offline processing to real-time \textit{big data} applications driven by the Internet of Things (IoT), robotics and autonomous agents.

This combination of factors has led to a sudden exploration of alternative computing methodologies that span the entire Boolean computational stack from physical effects, to materials, devices, architectures, and data representations. It also includes novel non-Boolean methods of computing such as quantum, wave and neuromorphic computation, Boltzmann machines and others. Exactly which combination of computational elements will evolve from this plethora of options is far from clear. However, it is possible to state general requirements future computing platforms must meet. First, any new computing methodology must be compatible with the existing multi-trillion dollar infrastructure associated with current CMOS based computing. Second, it must be scalable through multiple generations of incremental hardware and software improvements and, third, the performance/cost metric must exceed that of Boolean CMOS processors. 

One approach stems from the intuition that physical systems, governed by their underlying physical laws, effectively compute the evolution of their state in a natural way. More abstractly, matter implicitly stores the correlated physical information defining its dynamical state. It feeds back this information through its interacting environment such that both may modulate and respond to each other's properties. If an experimentally observable physical system can be constructed whose spontaneous dynamics map into solutions of a given problem, we can state that the system's matter has been used for computations. This paper will outline a specific physical system of such type. By combining the use of electron transport through random magnetic textures~\cite{Prychynenko2017,Bourianoff2018} and the principles of reservoir computing~\cite{Jaeger2007}, we will demonstrate the possibility of using magnetic systems as analog computers capable of processing and classifying data by means of sensitivity to its correlation features~\cite{Tanaka2019}. 

\begin{figure*}[t]
	\centerline{\includegraphics[width=0.8\textwidth]{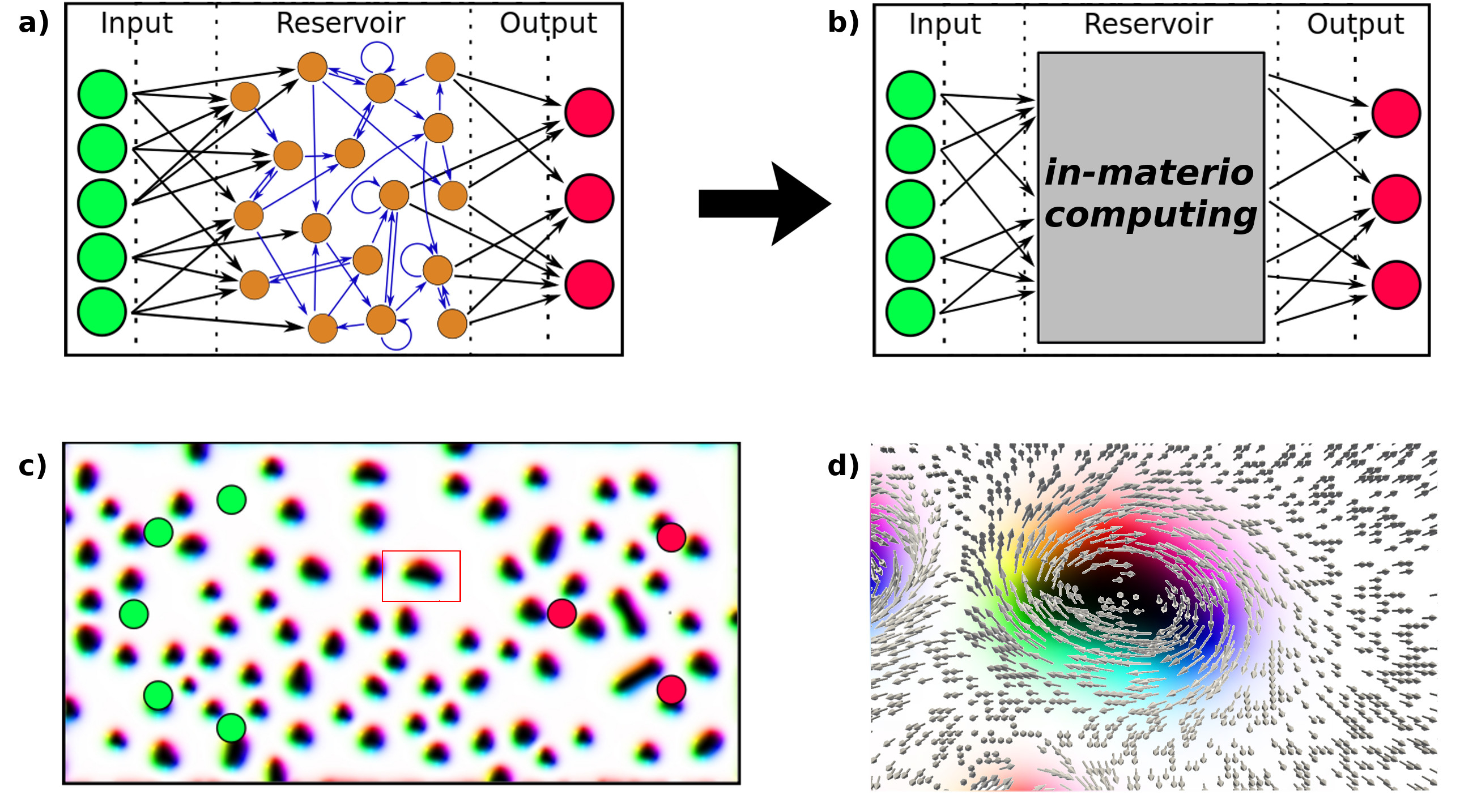}}
	\caption{{\footnotesize General schematic of reservoir computing for (a) a recurrent neural network and (b) a material-based reservoir. (c) Example of a skyrmion fabric reservoir with locations of input (output) contacts identified by green (red) dots. The in-plane orientation of the magnetization is color coded and (d) shown in detail with arrows for a sample Bloch skyrmion identified by the red frame in (c). The spatially extended nature of such a magnetic system allows for tunability in the number of electrical contacts. This allows to control the dimensionality of the reservoir snapshot.}}
	\label{fig:1_RCschema}
\end{figure*}

The remainder of the paper is structured as follows. To stress the importance and generality of the principles espoused by our results, in Section~\ref{sec:Neuro} we briefly review the salient properties allowing physical material systems to intrinsically compute. Starting with the success of deep neural networks (DNNs), we expand to recurrent neural networks (RNNs) in an effort to properly contextualize and introduce the concepts of Reservoir Computing (RC) in a contained, yet self-consistent way. Section~\ref{sec:model} will then introduce our physical model in detail and argue how the interplay of sensitivity to driving currents, magnetoresistance and intrinsic memory effects can be harnessed for RC applications. We do so by discussing the importance that the delay between current density and magnetic relaxation plays on defining memory in a magnetic system. Section~\ref{sec:results} will outline the numerical simulation of the models presented and verify both the non-linear processing performed by the skyrmion fabric on sample input signals as well as confirm the role played by the different natural timescales on the emergence of memory in the system. We proceed to use these tools to demonstrate pattern recognition of random pulse sequences using both time and spatially multiplexed techniques to elucidate the importance of tunable dimensionality in magnetic RC. In conclusion, Section~\ref{sec:conclusion} will further discuss the flexibility of our proposed design and reiterate the attention that should be directed towards exploring novel reservoir systems in nature.

\section{Neuromorphic Computing Principles}
\label{sec:Neuro}
\textbf{Deep Neural Networks --} The most proficiently used machine learning technique geared towards heavy data processing and pattern recognition is the DNN framework~\cite{Dietterich2002, Kotsiantis2007, Nasrabadi2007}. This consists of simulated weighted graph networks where vertex values ascribed to each node (i.e. neuron) are sequentially updated via a normalized weighted sum of the node values feeding into it. These nodes can in turn be used to emulate any arbitrary vector function of some input subject to the edge weights being chosen accordingly. This process, known as training, effectively reduces to finding the minimizer of a complex cost function. DNNs have enjoyed wide acclaim for their ability at outperforming humans in tasks previously considered unattainable by machines, going as far as beating the human world champion of the game `Go'~\cite{Silver2016}. The most general class of DNNs, known as \textit{recurrent} neural networks (RNNs) (see Fig.~\ref{fig:1_RCschema}a), considers coupled neurons whose graph representation consists of cycles forcing information to feedback (or \textit{echo}) throughout the network and fade over time~\cite{Pineda1987, Williams1989, Schmidhuber1992}. This basic property allows RNNs to exhibit complex temporal dynamic behavior in response to the interplay of the instantaneous driving inputs and the RNN's implicit internal memory's echoing of past input information. Whereas the universal approximation theorem for feed-forward DNNs guarantees the faithful representation of any function~\cite{Cybenko1989, Hornik1991}, the RNN's capacity to harness temporal correlations through their echo-state memories allows them to emulate universal Turing machines~\cite{Dale2017}. This has far-reaching consequences as RNNs can be considered universal approximators of dynamical systems as a whole.

What has not received much attention are the pitfalls that such algorithms typically suffer. The topology of neural networks is largely restricted by the ability to train the large number of weights. RNNs are particularly penalized as their training converges much more slowly (if at all) than simpler feed-forward networks due to the \textit{vanishing gradient problem}~\cite{Hochreiter1998, Hochreiter2009}. Approaches to overcome this fundamental difficulty have been proposed in the form of custom modifications to the general RNN structure~\cite{Hochreiter1997, Gers2000} - known as \textit{long short-term memory} (LSTM) networks - with widely recognized success in tasks such as handwriting~\cite{Graves2009} and speech~\cite{Graves2013} recognition. Furthermore, LSTM networks, have allowed the use of RNNs to enter mainstream use for specific voice recognition tasks on handheld devices~\cite{Graves2014, Sak2015, Wu2016, Thanda2016, Gehring2017, Xiong2018}. Nonetheless, these state-of-the-art algorithms still require the use of general purpose computers, thus inheriting all the end-of-Moore-law scaling bottlenecks~\cite{Waldrop2016}. Special purpose integrated circuit implementations such as IBM's {\it TrueNorth} chip, with its $65\,n\mathrm{W}/\mathrm{neuron}$ power consumption~\cite{Akopyan2015}, pale with the cerebral cortex's $0.2\,n\mathrm{W}/\mathrm{neuron}$. Deposition techniques used in the fabrication of microelectronic devices constrain special purpose hardware devices by imposing heavy challenges to scaling topologies requiring many-to-many connections. As an example, compare {\it TrueNorth}'s 256 connection fan in/out versus the brain's $\sim 10^4$. Furthermore, clock/power distribution and data interconnection have significant impact on overall chip power dissipation and complexity. Last but not least, successfully trained DNNs still exhibit impractically long inference times ($10^{-3}-1\,\mathrm{s}$~\cite{lin2014microsoft,veit2016coco}), defined as the time between input submission and output reception. This requires that data must be processed \textit{off\-line}, precluding applications requiring the ability to process large amounts of real-time data. Ultimately, these bottlenecks arise from the need to model, store and especially train the many synaptic weights and node activation functions across the numerous hidden layers in RNNs.

\textbf{Reservoir Computing --} Early works however noticed that Hebbian training of RNNs mostly modifies just the output weights linking the bulk of the network to the read-out layer~\cite{Dominey1995, Buonomano1995, Schiller2005, Dominey2006}. These were the first to explicitly suggest the principle of employing a - mostly - randomly weighted RNN. The claim is that a sufficiently large RNN can be initialized with static random weights. Its training is then solely performed on a smaller set of weights associated with feed-forward connections between the bulk of the network and the output layer~\cite{Jaeger2001, Jaeger2001a, Jaeger2007, Maass2002, Maass2011}. Since such output weights consist of only a single layer connecting to the static RNN bulk (known as the \textit{reservoir}), the training reduces to a linear regression performed on the reservoir state to satisfy a small training set of data. The advantage of this method, known as Reservoir Computing (RC), is not earned for free as the reservoir network has to typically be topologically more complex than an equivalent RNN fully trained to perform an identical task. 

\begin{figure}[t]
	\centerline{\includegraphics[width=3.2in]{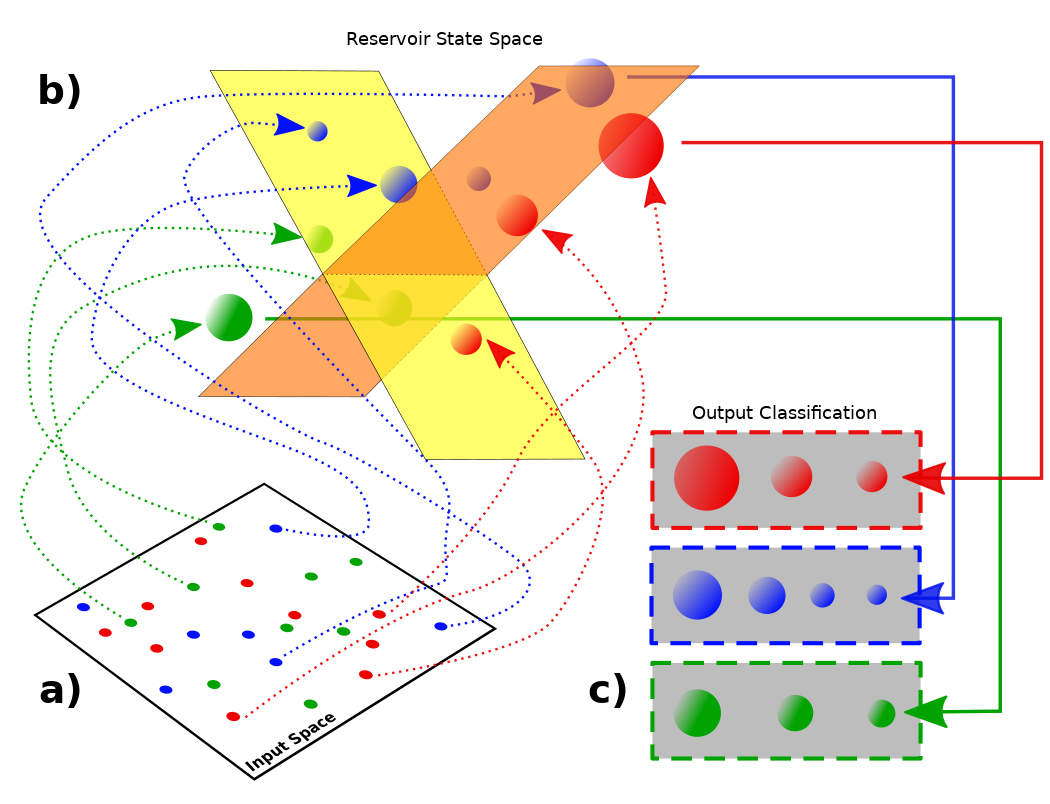}}
	\caption{{\footnotesize Schematic of the reservoir computing operational principle. (a) Unstructured data from an input space is (b) non-linearly projected by the reservoir's transient dynamics onto its higher dimensional state space. Due to the similar evolution of the reservoir when driven similarly correlated input data samples (represented by similar colors), (c) a single linear regression step can be used to define hyper-planes in the reservoir's state space such that different input data categories become separated. The task of the reservoir is to project different spatial-temporal events onto a sparsely populated high dimensional space where they become easier to recognize and categorize.}}
	\label{fig:2_RCflow}
\end{figure}

The principles supporting the feasibility of this scheme lie in the purpose served by the large, static reservoir. Tasks such as pattern classification require the neural network to emphasize the data separation of, initially ambiguous, input patterns. A highly connected reservoir performs a high dimensional nonlinear projection of the input data onto a large state manifold where the partitioning of the projected states is simplified via the introduction of hyper-planes (see Fig.~\ref{fig:2_RCflow}) defined by the output layer (see Appendix A). The non-linearity of this transformation, coupled with the RNN's memory, will guarantee that the reservoir's response to a given input will be sensitive to spatial-temporal correlations present in the input data commensurable to the memory's timescale. Distinct, but similarly correlated, input patterns will induce an evolution of the reservoir along similar trajectories. Each trajectory bundle (or \textit{phase tube}) corresponding to different input categories, rapidly separates from the others. The trained feed-forward output layer then partitions the reservoir's manifold via hyper-planes which unambiguously separate each phase tube. Since the network's reservoir is never modified, a sufficiently complex reservoir can in principle process different data-sets independently of how they are structured~\cite{Legenstein2005}. Each data-set will require solely the training of a new, dedicated, output layer. The reservoir also provides an ideal medium for sensor fusion with multiple sensors may feed into a common reservoir and the output layer trained to recognize multi sensor events e.g.\ audio-visual events.

Whereas RC offers to significantly simplify the training complexities of RNNs, the need for a potentially much larger network reservoir typically exacerbates the computational resource and inference time requirements needed for successful modeling. However, since the reservoir is only a static complex nonlinear dynamical system, it is not necessary for it to be implemented in the form of a RNN. Any sufficiently nonlinear and high dimensional analog physical system can in principle perform the same role~\cite{schrauwen2007overview} (see Fig.~\ref{fig:1_RCschema}b). This paradigmatically different approach to computing - known as \textit{in-materio computing} - seeks to overcome the digital limits of DNNs by leveraging the vast amounts of readily accessible physical complexity already present in nature (for a review see~\cite{Dale2017}). Following initial demonstrations involving single non-linear dynamical nodes with delayed feedback~\cite{Appeltant2011}, recent work has explored RC with physical systems such as liquid crystals~\cite{Miller2002}, water buckets~\cite{Fernando2003}, bacterial colonies~\cite{Jones2007}, memristors~\cite{Kulkarni2012}, optical resonators~\cite{Duport2012, Larger2012, Brunner2013, Duport2016} and poly-butyl-methacrylate mixtures~\cite{Mohid2014a,Mohid2014b,Mohid2014c}. In magnetic systems, single spin-vortex nano oscillators have been used to demonstrate spoken digit recognition within the RC framework~\cite{Torrejon2017} and skyrmions suggested for further RC development~\cite{Prychynenko2017,Bourianoff2018}. The reservoir computing paradigm provides scientists with a general, rigorous and powerful tool to tackle tasks that require temporal dependencies, without employing hand-crafted data representations. 

Generally speaking, a performant reservoir has the following properties:

\begin{enumerate}
 \item The dimension of the reservoir's phase space must be much larger than the size of the input category set.  
 \item To guarantee reproducibility of its dynamical responses to identical inputs, the reservoir has to relax or be resettable to the same initial state once all inputs are removed.
 \item To ensure that memory of features fully affects the reservoir's evolution, any temporal feature correlations present in the input data must be of the same order as the natural transient dynamical timescale of the reservoir.
 \item The dynamics of the reservoir must be nonlinear but not ergodic to the point of strongly mixing trajectories throughout reservoir evolution.
 \item To properly identify all desired feature categories, a sufficiently large enough subspace of the reservoir state should be measurable.
\end{enumerate}

The last requirement is particularly important in light of the dynamical systems interpretation used to justify reservoir computing. The feasibility of exploiting a physical system as a reservoir intimately depends on the technical ability to reliably inject input data and sample a sufficiently high dimensional subspace of the reservoir's state. To capture the reservoir's separation of $N$ phase tubes corresponding to the number of input data categories, the output layer should be capable of generating a sufficient number hyper-planes to distinguish each phase tube pair.

Since a properly functioning reservoir is input data agnostic, any device implementation must have the flexibility of tuning the dimensionality of the output layer to effectively classify diverse data sets. Many of the reservoir computing examples presented in the literature do not allow such tunability~\cite{Appeltant2011, Paquot2012, Martinenghi2012, Brunner2013, Torrejon2017}, preferring to  sample temporal traces of a low dimensional substate of the reservoir over time. This is also necessary whenever the input data's temporal correlations are widely different from the reservoir's natural timescales and need to be preprocessed to mix input data features together before exciting the reservoir: often a nonlinear transformation in itself. These time trace values are then used to artificially boost the dimensionality of the readout without actually exploiting the inherently high dimensional nature of the reservoir's state space. As a result, it is often unclear how much of the classification work has been performed by the reservoir as opposed to the nonlinear input data preparation and time trace measuring performed by the output layer~\cite{Torrejon2017}. An ideal reservoir design should allow for arbitrary tunability of the dimensionality of both input and output layers. This in turn allows for a true RC implementation where unstructured data may be injected without any preprocessing and the reservoir state sampled via instantaneous, real-time, snapshots. 

\textbf{Magnetic systems as Reservoirs --} There is no dearth of physical systems in nature which satisfy the general properties just discussed. Assessing, however, which systems may be industrially viable is an entirely different challenge. This work argues for the employment of magnetic textures due to their nanometer sizes, intricate dynamical properties and, most importantly, low-power and CMOS-compatible all-electrical operability. In particular, we propose using random topological magnetic textures -- skyrmion fabrics -- to generate complex, high-dimensional representations of input voltage signals (Fig.~\ref{fig:1_RCschema}c). Magnetic skyrmions are compact and meta-stable magnetic structures predicted over two decades ago~\cite{Bogdanov1989} and very actively studied experimentally both in lattice~\cite{Muhlbauer2009a} and isolated form~\cite{Romming2013} (see Fig.~\ref{fig:1_RCschema}d). The particle-like properties of skyrmions have been extensively summarized in several reviews~\cite{Nagaosa2013, Finocchio2016, Fert2017, Jiang2017, ES2018}. Their mobility under ultra-low current driving~\cite{Jonietz2010,Schulz2012} and room-temperature stability~\cite{Yu2011b, Yu2012,Gilbert2015, Boulle2016, Woo2016, Tomasello2017a} have garnered them a central position as information carriers in many device-relevant materials and applications~\cite{Fert2013, Tomasello2014, Zhang2015c, Muller2016, Pinna2018}. Device oriented research has however mostly ignored intermediate skyrmion phases -- known as ``skyrmion fabrics''\cite{Prychynenko2017, Bourianoff2018} -- which interpolate between single skyrmions, skyrmion crystals and magnetic domain walls\cite{You2015} (an example is shown in Fig.~\ref{fig:3_grains}b). We claim that the random phase structure, complexity, nonlinear response, and memory characteristics present in skyrmion fabrics justify their use as a reservoir for RC. Gigahertz voltage patterns exciting the texture via nanocontacts can implement the injection of input information while the coupling of electron transport and magnetoresistivity~\cite{McGuire1975, Hanneken2015, Kubetzka2017} can be used to sample the magnetic state of the system.

\section{Skyrmion Fabric Reservoir Model}
\label{sec:model}

\begin{figure}
\includegraphics[width=0.45\textwidth]{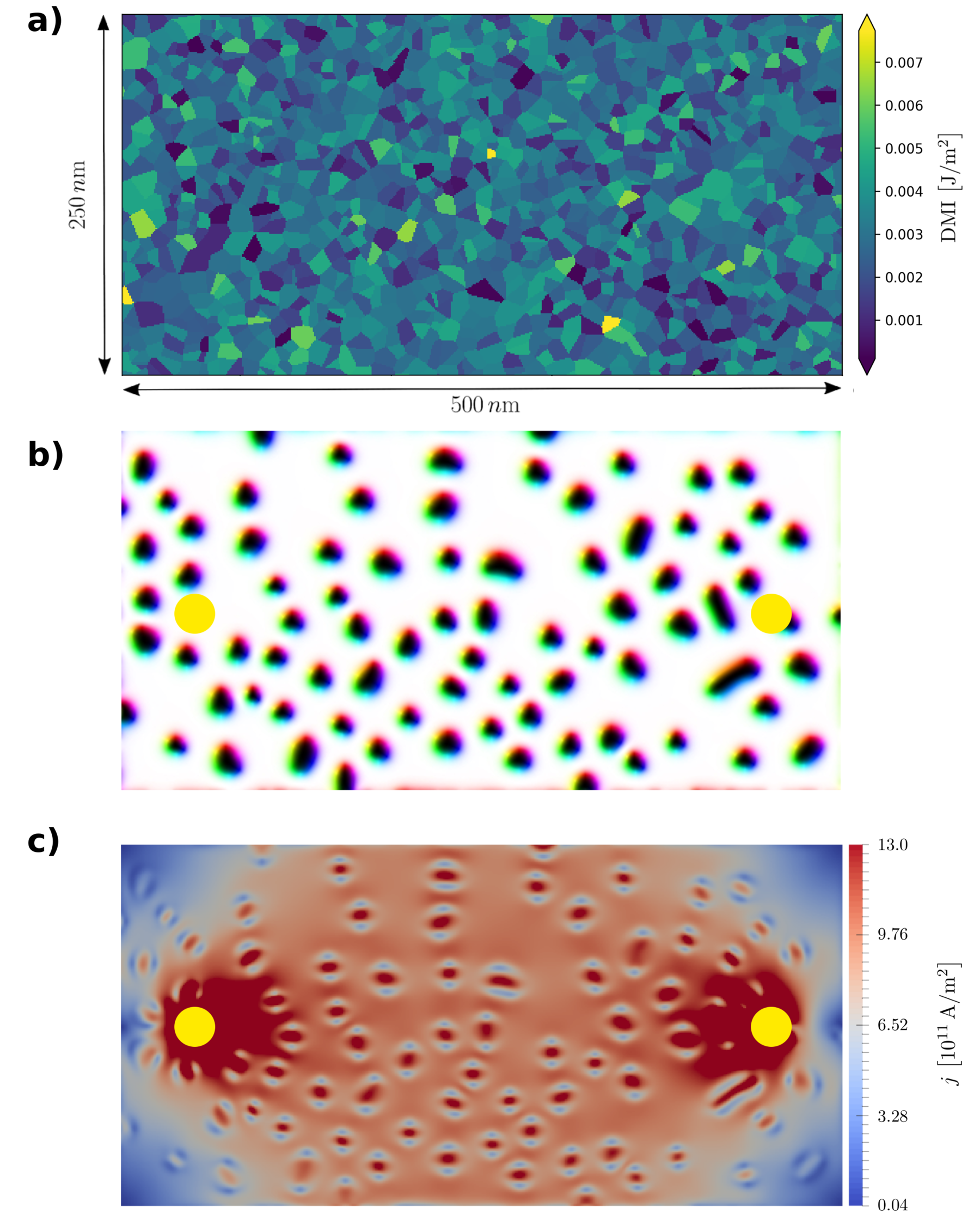}
\caption{(a) Randomly generated DMI-grain inhomogeneities employed to pin skyrmions. Figure shown corresponds to a 250-by-500 $n\mathrm{m}$ geometry consisting of $10\,n\mathrm{m}$ grains exhibiting a $40\%$ DMI variance around a $0.003 \mathrm{J}/\mathrm{m}^2$ mean value. (b) Sample skyrmion texture used for simulations. The locations of the electrical nanocontacts are identified by the yellow disks (size has been enhanced for visibility). (c) Sample relaxed current distribution through the texture shown in (b) when a $110\,m\mathrm{V}$ potential difference is applied across the nanocontacts.}
\label{fig:3_grains}
\end{figure}

\begin{figure}
\includegraphics[width=0.45\textwidth]{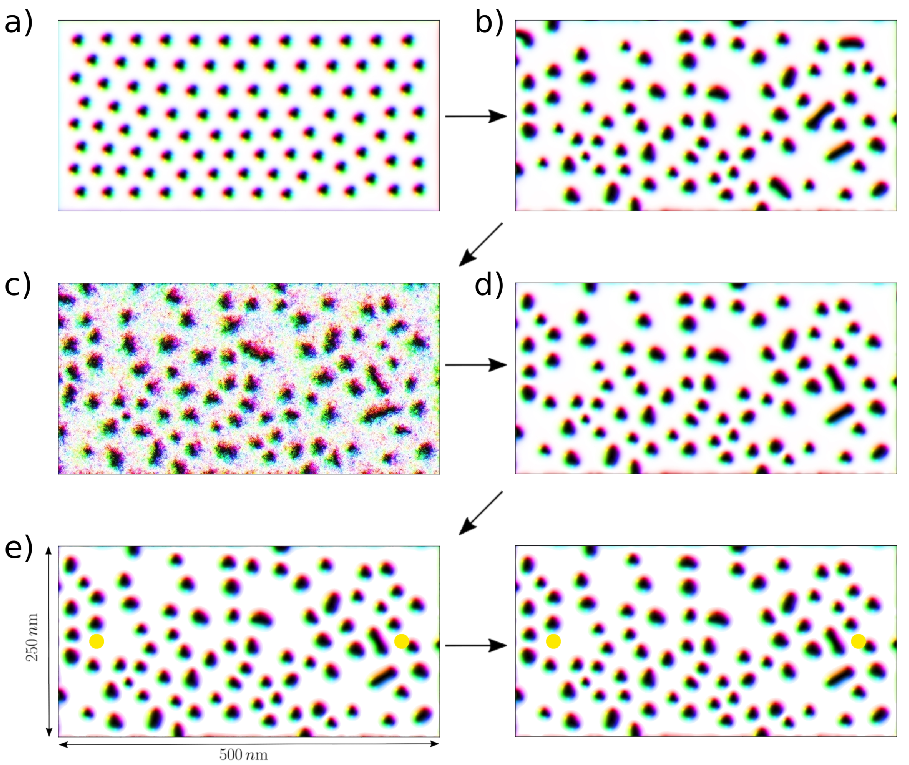}
\caption{Initialization of magnetic texture and thermal stability test. (a) An artificial skyrmion lattice is generated. (b) The lattice is relaxed under the effect of an applied external magnetic field and DMI grain inhomogeneities. (c) Thermal noise is added and allowed to act for $20\,n\mathrm{s}$ before being switched off. (d) The magnetic texture is relaxed again in the absence of thermal noise. (e) A $20\,n\mathrm{s}$ constant voltage pulse is applied to verify that skyrmions are not displaced by current-mediated transport effects. The relaxed magnetic configurations in (b) and (d) are compared to verify that the majority of skyrmions appearing in the bulk of the geometry are not significantly affected by thermal effects.}
\label{fig:4_thermtest}
\end{figure}

For the sake of simplicity, we consider a random magnetic skyrmion phase in a spatially extended rectangular geometry with only two voltage contacts. To model a realistic setup, our sample is subject to Dzyaloshinskii-Moriya interaction(DMI) via grain inhomogeneities and a static applied magnetic field (see Fig.~\ref{fig:3_grains} and Methods Section). A random magnetic texture is generated by imposing an initial skyrmion lattice structure and allowing it to freely relax (see Fig.~\ref{fig:4_thermtest}). The introduced magnetic inhomogeneities have been extensively studied in the literature as a source of skyrmion pinning to explain the discrete onset of the creep threshold in skyrmion mobility~\cite{ Legrand2017, Kim2017a} both theoretically and in experiments~\cite{Schulz2012}. Initialization is considered complete when the magnetic texture has relaxed to a stable state and does not change significantly when subject to thermal noise and a constant applied voltage across the electrical contacts. The voltage magnitude is chosen such that the resulting magnetization dynamics lie just below the skyrmion creep threshold where their deformations are maximal.

As aforementioned, the magnetic texture is excited via time-varying voltage patterns injected through the nanocontacts (refer to Sec~\ref{sec:methods} for details). Due to the sub-creep setup described, such patterns will excite time dependent deformations of the magnetic skyrmion texture due to a variety of magnetoresistive effects~\cite{McGuire1975, Hanneken2015, Kubetzka2017}. Since the natural electron relaxation timescale is orders of magnitude smaller than the ferromagnetic resonance (FMR) timescale ($\sim 10^{-14}\,\mathrm{s}$ vs. $\sim 10^{-9}\,\mathrm{s}$), these effects will guarantee that a given state of the magnetic texture will result in a unique corresponding current distribution throughout the geometry. To simplify the modeling of such electron-transport mediated effects and isolate their qualitative nature, we will focus solely on the anisotropic magnetoresistance (AMR) effect. Current densities are calculated self-consistently through $\mathbf{j}[U, \mathbf{m}] = -\sigma[\mathbf{m}] \cdot \mathbf{E}[U]$ at each time step of the magnetization's Landau-Lifshitz-Gilbert (LLG) dynamics~\cite{Kruger2011}. The electric field through the texture, induced by the applied voltage, is calculated by solving the Poisson equation $\mathbf{E}=-\nabla \Phi$ with boundary conditions \mbox{$\Phi|_{c1}=-\Phi|_{c2}=U$} at the two contacts, and the conductivity tensor $\sigma [\mathbf{m}] = \frac{1}{\rho_{\perp}} \mathds{1} + \left( \frac{1}{\rho_{\parallel}} - \frac{1}{\rho_{\perp}} \right) \mathbf{m} \otimes \mathbf{m}$ is computed at each point throughout the geometry. We denote by $\rho_{\perp}$ ($\rho_{\parallel}$) the current resistivities for flows perpendicular (parallel) to the magnetization direction.  For definiteness the results in the next section consider the typical case where $\rho_{\perp}>\rho_{\parallel}$. An example of an excited skyrmion fabric is shown in Fig.~\ref{fig:3_grains}b right above a density plot of the instantaneous current distribution traversing it (Fig.~\ref{fig:3_grains}c).

\begin{figure*}
	\centerline{\includegraphics[width=0.85 \textwidth]{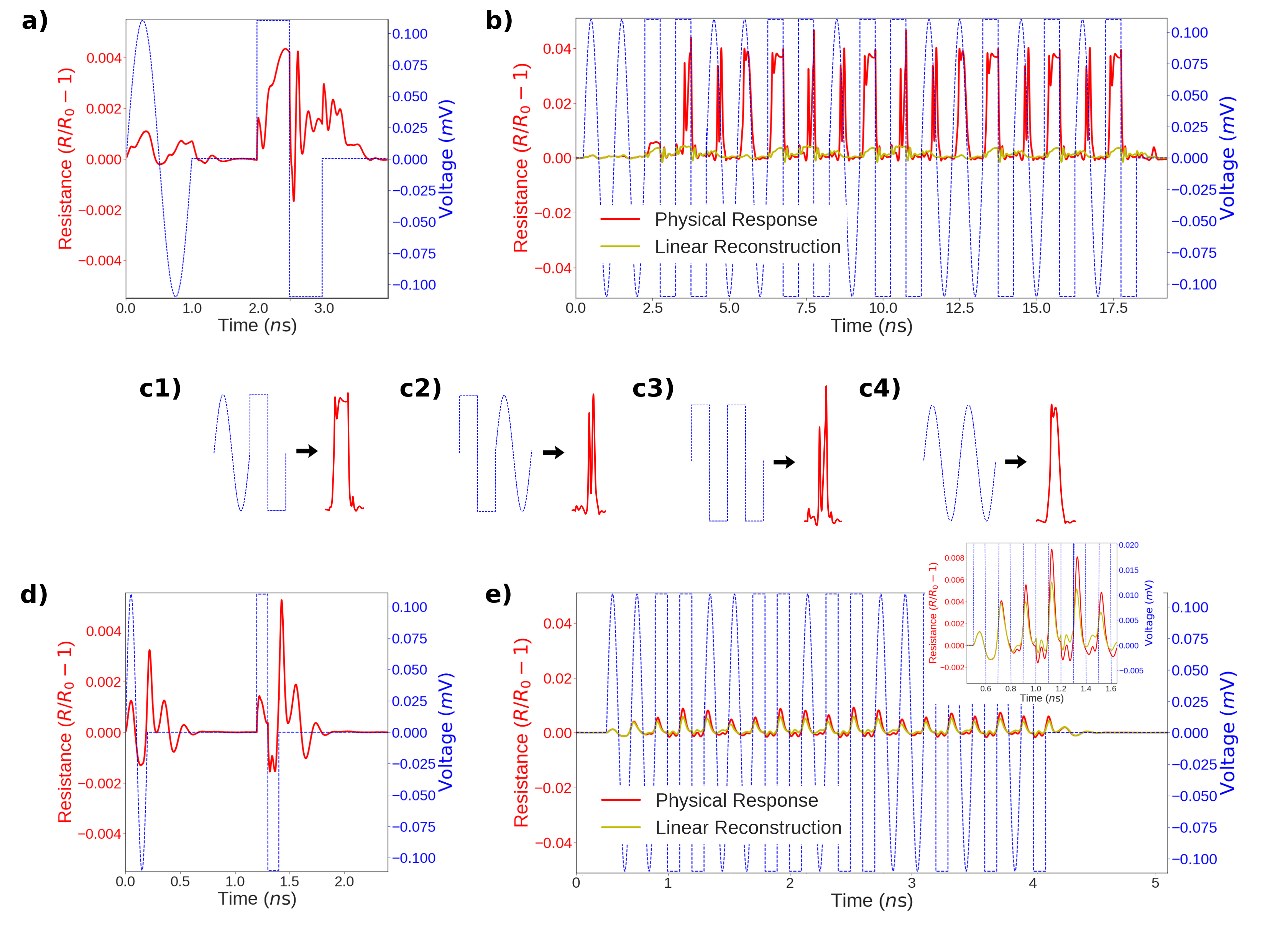}}
	\vspace{-3mm}
	\caption{{\footnotesize AMR response (red) of magnetic texture to individual sine and square voltage pulses (dashed blue) used to drive the magnetization dynamics with frequencies: (a) $1\,G\mathrm{Hz}$ and (d) $5\,G\mathrm{Hz}$. AMR response to identically structured pulse trains used to drive the magnetic texture with varying combinations of sine and square voltage pulses with frequencies: (b) $1\,G\mathrm{Hz}$ and (e) $5\,G\mathrm{Hz}$. Resistances are given in units of $R_0$, which denotes the geometry's AMR resistance in the vanishing applied voltage limit. In (a) and (d), the ring-down of the magnetic texture is exemplified by the delay between the voltage pulse ending and the AMR relaxation to equilibrium. In (b) and (e), the AMR response observed for individual pulses is superimposed consistently with the driving voltage (yellow) for comparison to the simulated AMR response to emphasize its nonlinear behavior. At frequencies much larger than the system's ferromagnetic resonant (FMR) frequency (e) the magnetic texture's nonlinear evolution does not have sufficient time to respond to the driving voltage patterns leading to a close match of both the AMR response of (d) the individual pulses as well as the (e) response of the pulse train with its linear reconstruction (see inset for details). At frequencies comparable with the FMR frequency the magnetization dynamics are highly sensitive to both the instantaneous voltage intensity and memory of past voltage values, presenting a large spike in peak AMR resistance. This is apparent in both the AMR response to (a) individual sine and square pulses as well as (b) a strong difference between the pulse train's response and the linear reconstruction. The AMR response observed in (b) is characterized by four reproducible and distinct resistance pulse shapes depending on whether a (c1) sine-square, (c2) square-sine, (c3) square-square or a (c4) sine-sine pulse combination is driving the dynamics.}}
	\label{fig:PulseTrain}
\end{figure*}

Previous work\cite{Prychynenko2017} has focused on details of current paths in the presence of single magnetic Bloch and N\'eel skyrmions as well as current density distributions through static random textures\cite{Bourianoff2018}. Skyrmions have been shown to exhibit non-linear current-voltage characteristics due to the interplay of magnetoresitive effects and pinning. Furthermore, the highly irregular current distribution imposed by the complex magnetic texture does not simply converge to a single localized path of least resistance across the contacts. It instead distributes throughout the magnetic texture, thus interacting with the entire geometry as a whole (as seen in Fig.~\ref{fig:3_grains}c).  We leverage these properties to explore how stable random magnetic textures can non-linearly process time varying voltage signals. The AMR-mediated current flow, as observed through the net contact-to-contact resistance, allows for the verification that the complex magnetization dynamics truly process voltage signals in such a way that their temporal-correlation features are preserved.

\section{Results}
\label{sec:results}

We initially test the texture's AMR response to individual sinusoidal and square pulses of varying frequencies ranging from $0.7\,G\mathrm{Hz}$ to $5\,G\mathrm{Hz}$ with $1\,n\mathrm{s}$ `resting' intervals between them (see Appendix~\ref{app:B} for supplemental data). This allows us to both further test the reproducibility of the AMR response as well as observe the lag in the magnetization dynamics relative to the voltage driving. The magnetic texture we consider, modeled after recently studied Pt/Co/Ir multilayer stacks (see Methods Section for system details), is seen to remain in a dynamical state past the end of the voltage driving, requiring and extra $\sim 0.6 n\mathrm{s}$ to settle back down to its equilibrium state (see Figs.~\ref{fig:PulseTrain}a,~\ref{fig:PulseTrain}d). The AMR resistance response to individual pulses is found to be quite modest, never exceeding $0.5$\% at any of the frequencies considered.

We then proceeded to continuously excite the texture with random sequences of square and sinusoidal pulses and observe their resulting AMR responses (see Figs.~\ref{fig:PulseTrain}b,~\ref{fig:PulseTrain}e). To quantify how much the resistance trace measurement can be considered a nonlinear processing of the driving, we consider a hypothetical resistance response constructed by linearly superposing the individual pulse responses discussed above. We find that at high driving frequencies (Fig.~\ref{fig:PulseTrain}e) the observed AMR response does not differ significantly from the linearly constructed response. As the driving frequencies are lowered (Fig.~\ref{fig:PulseTrain}b), the differences between the two grow larger denoting a strong non-linear filtering effect resulting from the magnetization's out-of-equilibrium dynamics having time to fully interact with the variations in driving voltage. We attribute this dependence to the relative scale of the voltage driving and magnetic texture's FMR frequencies. As the FMR timescale effectively denotes the time taken by the magnetization to respond to its effective field, the magnetization dynamics are maximally susceptible to driving effects on or close to this timescale. This is further exemplified by the sudden increase in the AMR response amplitude (to values as high as $4$\%) as the sampled frequencies are lowered down into the $0.7-1.8\,G\mathrm{Hz}$ range (see Appendix A for more frequency data).

Focusing on the $1\,G\mathrm{Hz}$ AMR response data shown in Fig.~\ref{fig:PulseTrain}b, we note that the resistance trace consists of four reproducible resistance pulse shapes which repeat throughout the driving (highlighted in Fig.~\ref{fig:PulseTrain}c). These resistance traces can be seen to carry information about the voltage signal's history. The type of resistance response pulse observed can be seen by plain inspection to match not just with the shape of the instantaneous voltage pulse driving the magnetic texture but also the voltage pulse immediately preceding it. As the voltage pulse train consists only of sine and square waves, we have four possible two-pulse combinations: sine-square (c1), square-sine (c2), square-square (c3) and sine-sine (c4). We argue for this to be strong evidence of memory effects in the magnetic texture for temporally correlated features. The lagged response time of the magnetic texture to a driving pulse implies that the magnetization can be expected to continue carrying information of its driving history. As such, any new pulses applied within the magnetization memory window will lead to a response which depends not just on the pulse applied but also on past pulses still naturally present in the memory of the system.

\begin{figure}
	\centerline{\includegraphics[width=0.5 \textwidth]{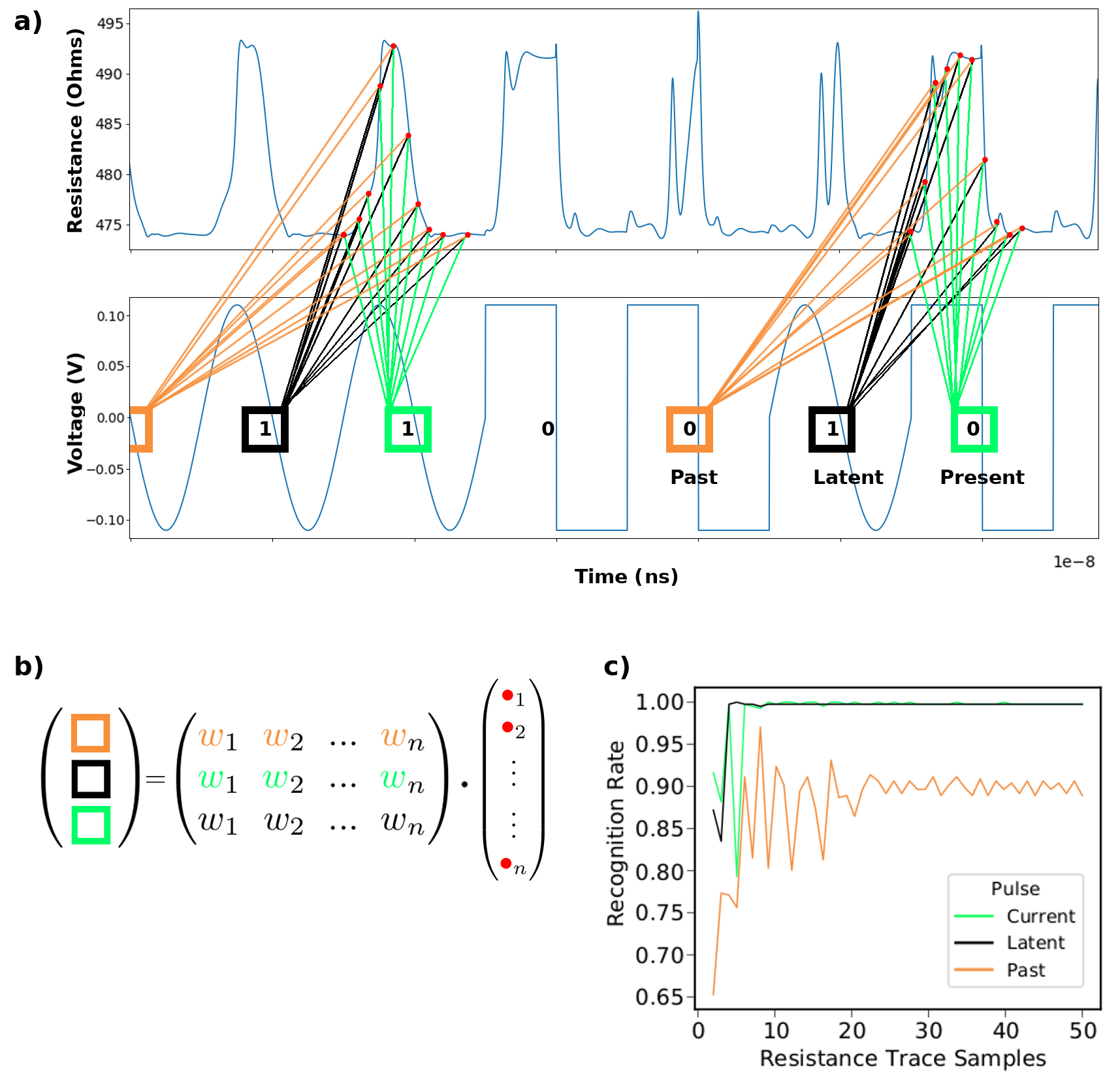}}
	\caption{{\footnotesize a) Schematic of the time tracing used to train and operate the output layer (temporal axis is common for both voltage and resistance plots). The resistance response is sampled at regular intervals across the temporal window of the instantaneous - {\it current} - pulse driving the magnetic texture. b) These resistance values are then summed in weight for three different sets of weights. One set is trained to recognize the {\it present} pulse  currently driving the system, the second set is trained to recognize the {\it latent} pulse preceding the {\it present} one, and the third set is trained to recognize the {\it past} pulse preceding the {\it latent} one. c) Recognition rates of {\it present}, {\it latent} and {\it past} pulses as a function of the number of resistance trace samples chosen.}}
	\label{fig:timeplex}
\end{figure}

\textbf{Pattern Recognition --} To more precisely quantify these concepts we trained a single feed-forward output layer to attempt recognizing all possible three-pulse sequences ($2^3=8$ in total) by simulating the driven dynamics of our skyrmion texture subject to a $400\,n\mathrm{s}$-long random pulse train (refer to Appendix~\ref{app:A} for extended details). For sake of nomenclature, we will refer to the {\it present pulse} as the current pulse driving the texture concurrently to its state probing, {\it latent pulse} as that which preceded the present pulse, and {\it past pulse} as that preceding the latent pulse (see Figure~\ref{fig:timeplex}). To exemplify the reservoir computing properties, the training set consisted of only one example of each 3-pulse sequence and sampling of our reservoir state was performed in two different ways to compare memory performance. 

For the first technique, we employ time-multiplexing to sample the time-resolved AMR response at regular intervals across the time window corresponding to the present pulse driving. This results in a vector of resistance values of dimension fixed by the number of samples chosen. As all these resistance values are a function of the magnetic state, only throughout the present pulse driving, any of their capacity to recognize latent and past pulses can only be the result of memory in the present magnetic state. As schematically shown in Fig.~\ref{fig:timeplex}a, these resistance values are summed in weight and to generate a recognition of present, latent and past pulses driving the system where the weights are trained via a standard linear classifier on one instance of each three-pulse combination. This linear classification is the realization of the hyperplane concept used to distinguish reservoir trajectories as described in the RC introduction part. As a function of the number of resistance samples, Fig.~\ref{fig:timeplex}b demonstrates the success of this method at recognizing patterns as far back as the past pulse. As all sampled resistance information takes place at more than $1\,n\mathrm{s}$ from the end of the past pulse, this implies that memory persists longer even than the magnetic relaxation timescale when the texture is out of equilibrium. These recognition rates are valid down to $\sim 10$ resistance samples showing how this technique could potentially be viable as long as the resistance can be time-resolved down to the $\sim 0.1\,n\mathrm{s}$ timescale.

\begin{figure}
	\centerline{\includegraphics[width=0.55 \textwidth]{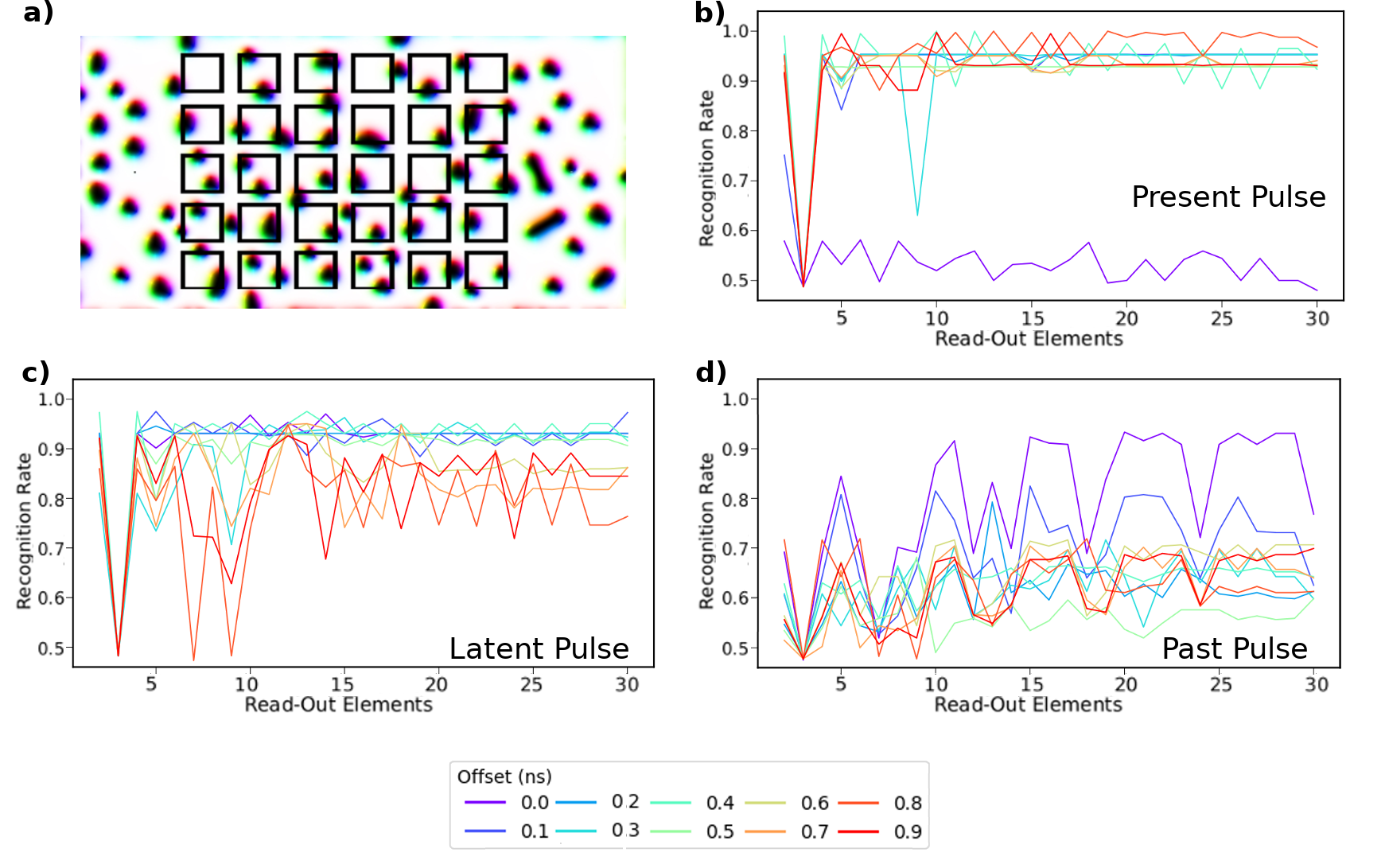}}
	\caption{{\footnotesize a) Schematic of read-out element grid used to measure $20 n\mathrm{m}^2$ averages of the out-of-plane component of the magnetic texture. Recognition rate of the voltage pulses b) present, c) latent, and d) past to the read-out measurement (see text and Fig.~\ref{fig:timeplex} for definition) as a function of the number of grid elements used. Color coding corresponds to offset time in nanoseconds from the beginning of the current pulse. At $0\,n\mathrm{s}$ offset, the recognition rate of the current pulse is $50\%$ as no information of the voltage pulse is actually present in the magnetic texture.}}
	\label{fig:spaceplex}
\end{figure}

To avoid these time resolution constraints altogether, we propose a second technique where the magnetic state is sampled spatially directly on the magnetic texture. We model a regular $5\times 6$ grid of $20 n\mathrm{m}^2$ regions separated $20 n\mathrm{m}$ from their neighbors (see schematic in Fig.~\ref{fig:spaceplex}a). In each region we compute the average out-of-plane component of the magnetization at a specific instant during the current pulse driving. Such an approach would be analogous to deploying a grid of magnetic tunnel junctions across the sample to capture a coarse snapshot of the magnetic state. As done for the time-multiplexing case, this array of average magnetization values is trained via linear classifier on our minimal 8-pulse training set and used to recognize the current, latent, and past pulses of our entire $400\,n\mathrm{s}$ data-set.

In Fig.~\ref{fig:spaceplex}b-c we show the pattern recognition results as a function of the number of read-out elements used (see \ref{sec:methods} for details) and color coded according to the time offset of their snapshot as measured from the beginning of the current pulse.  In Fig.~\ref{fig:spaceplex}b we see immediately how taking a snapshot right at the beginning of the current pulse results in a $50\%$ recognition rate of the current pulse itself. This is understandable as no information of the current pulse exists at this time and proper recognition is effectively equivalent to a coin toss. In Fig.~\ref{fig:spaceplex}c, we see how the recognition rates for the latent pulse prefer small offsets. This is consistent with the idea that information is lost over time due to dissipative effects. Finally, in Fig.~\ref{fig:spaceplex}d we see how the recognition rate drops dramatically for the past pulse as a function of increasing offset. The probabilities still remain appreciably above $50\%$ denoting some correlation between the transient magnetic state as opposed to the no-offset recognition rates of the current pulse at $0 n\mathrm{s}$ offset.

\begin{figure}
	\centerline{\includegraphics[width=0.5 \textwidth]{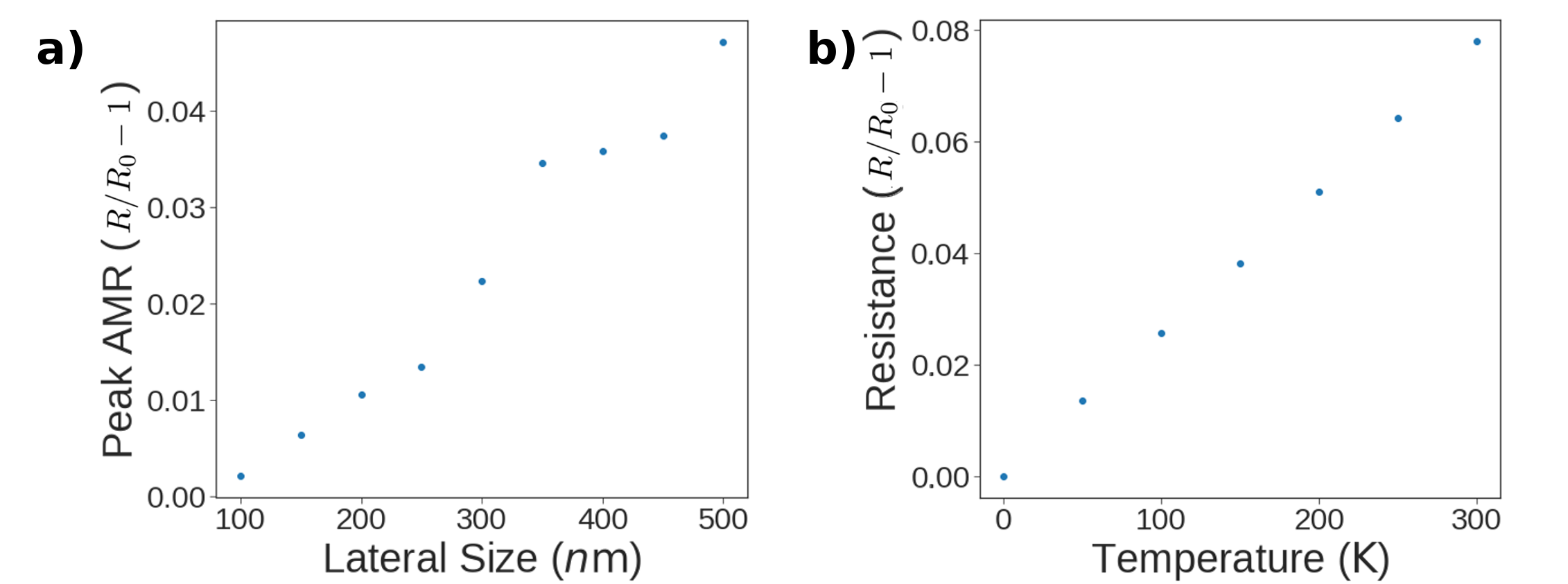}}
	\caption{{\footnotesize (a) Example peak AMR response to a $1\,G\mathrm{Hz}$ pulse train driving (see Fig.~\ref{fig:PulseTrain}) as a function of lateral size of the magnetic geometry (vertical size is kept constant). Given that material parameters and external fields effectively set the skyrmion density, peak resistance effects are expected to scale with the area of the magnetic geometry. (b) Average equilibrium AMR resistance of a $250\times 500\,n\mathrm{m}^2$ geometry as a function of ambient temperature in the limit of vanishing applied voltage. As in previous results, resistances are given in units of $R_0$, which denotes the zero temperature equilibrium of the $250\times 500\,n\mathrm{m}^2$ geometry. Combining results (a) and (b), we expect our results to be qualitatively valid for temperatures below $ 100\,\mathrm{K}$. A more quantitative understanding of these dependancies can only be gleaned through full electron transport simulations through the magnetic texture. Results discussed in Fig.~\ref{fig:PulseTrain} are expected to be qualitatively valid for temperatures under $100\,\mathrm{K}$.}}
	\label{fig:ResistanceTuning}
\end{figure}

To systematically study how the peak AMR amplitude scales with system size, we generated random textures on $250\times L\,n\mathrm{m}^2$ geometries where the lateral dimension $L$ was allowed to vary between $50-500\,n\mathrm{m}$ to modulate the net number of skyrmions participating in the signal-filtering process. The voltage applied at the contacts was scaled by the geometry's area to properly compare the current densities across the different simulated samples. Since all other physical parameters are kept identical, altering the size only changes the net number of skyrmions without altering their density and individual sizes. As expected, the AMR response (Fig.~\ref{fig:ResistanceTuning}a) is seen to mostly scale linearly with the lateral size. This implies that its magnitude is controlled by the total number of skyrmions deforming under the voltage driving. Furthermore, the AMR effect's intensity can be modulated by considering larger skyrmion fabric geometries.

Aside from the stability tests discussed in Fig.~\ref{fig:4_thermtest}, all results shown have been obtained via zero temperature micromagnetic simulations. Effects due to thermal noise can significantly affect the magnetic texture's dynamics and, consequently, the measured resistance responses. As full simulations of current transport {\it and} thermal effects are impractical due to time constraints given our current computational tools, we only show the sample's equilibrium AMR as a function of thermal noise to justify a lower bound on the signal-to-noise ratio that can be expected in such geometries (see Fig.~\ref{fig:ResistanceTuning}b). The equilibrium AMR (in units of the zero temperature magnitude $R_{0}\equiv R\vert_{\mathrm{T}=0\mathrm{K}}$) is seen to scale linearly with temperature up to  to a room temperature value of $\sim 8\%$. Since the peak AMR measurements of our sample when excited by $1\,G\mathrm{Hz}$ continuous voltage patterns was registered at $\sim 4\%$, we estimate that the results simulated in our geometries should be reproducible at temperatures below $100\,K$ for samples with similar geometries.

\section{Discussion and Conclusion}
\label{sec:conclusion}

In this work we have demonstrated how magnetic skyrmion fabrics can be employed for RC applications. The nonlinear character of the underlying magnetization dynamics subject to voltage pulses has been shown to exhibit intrinsic memory properties useful for identifying feature correlations of input driving voltages. The combination of non-linear effects, memory property and the very high dimensionality of a magnetic fabric's state space can guarantee enough separation of input data for the purpose of transforming classification tasks into linearly solvable ones. Furthermore, the $G\mathrm{Hz}$ speeds exhibited by such textures offer a large margin of improvement for such classification operations compared to the inference times of cutting edge DNN techniques ($\ge 10\,m\mathrm{s}$). By contrast, a magnetic reservoir such as the one described could reduce inference times by a factor of $1000$. This is especially so if the RC device is designed to not require complex preprocessing of inputs as regularly done in many RC schemes. A natural extension for even faster processing could eventually be offered by anti-ferromagnetic materials where the magnetization dynamics evolve up to $T\mathrm{Hz}$ speeds.

Whereas the time-multiplexing techniques used for the pattern recognition show greater recognition accuracy, it is in the space-multiplexed results and more so, in the combination of both, that we find the most promise. One can in fact think about combining only two snapshots of the texture over the span of one driving pulse to recognize current and latent pulses with fairly high accuracy. space-multiplexing allows for a tradeoff of time for spatial resolution which may be preferable for certain device designs. Furthermore, the time-multiplexing technique described will likely be strongly susceptible to thermal effects. Our current zero temperature implementation allows for a very deterministic propagation of information through the dynamics of the system. The minimal contribution of the past pulse to the current magnetic state, which allows the time-multiplexing to capture and use for recognition, is likely to be washed out by any randomizing effects. In both cases, we would like to emphasize that the training was performed on a very limited set of data consisting of only one example per three-pulse sequence (eight in total). Minimum training set size can be expected to grow as a function of the complexity of pattern recognition required and potential noise in the system. Regardless, RC schemes allow for a drastic reduction in training efforts by design due to the dimensionalities involved in the recognition process. While a more rigorous demonstration would also require introducing noise in the applied voltage pulses (currently beyond our computational capacity), we note that our present results demonstrate memory and absence of chaotic behavior in the underlying physical processes, thus justifying the usage of complex, spatially extended, magnetic textures for RC.

We expect for our setup to be experimentally realizable with current laboratory techniques, allowing for the exploration of multi-contact arrangements discussed in this work but currently beyond the practical limits of numerical simulation. This work has only considered the AMR contribution to the physics as a magnetoresistive effect but, realistically, many more effects may play a role. Among these, the non-collinear magnetoresistance has been even suggested as a reliable all-electrical detection scheme for magnetic skyrmions~\cite{Hanneken2015}. The skyrmion-based RC system utilizes the electron transport through the texture's domain walls embedded in the fabric in place of the top down interconnected structure utilized in conventional integrated circuits. This bypasses the fabrication and performance issues associated with nanoscale interconnects. It furthermore increases energy efficiency by avoiding the need to transduce magnetic information into electrical signals. Predicting the ultimate performance/cost metric for device applications is clearly impossible at this early stage but skyrmion based fabrics offer an attractive option for RC based on the results presented. As an estimate of power dissipation of the magnetic texture alone, given the applied voltage intensities and observed resistances, one modeled on this work would consume $1-100\,\mu\mathrm{W}$ of power depending on the magnetoresistive ratio of the material.

More broadly, these results represent a basic proof-of-concept for the foundational role that magnetism and spintronics can play in the development of such technologies. The general principles elucidated in the introduction can in fact be seen to apply to a wide category of physical systems, potentially spurring more interest in leveraging the gratuitous complexity that abounds in them. It is our opinion that this work should be interpreted as a wider call to action for further exploration of these endeavors by the condensed matter physics community. 

\section{Methods}
\label{sec:methods}

The micromagnetic simulations were performed using the MuMax3 GPU-accelerated micromagnetic simulation program~\cite{vansteenkiste2014design} for fast magnetization relaxation and the Micromagnum simulation program~\cite{selke2014design} with custom AMR module~\cite{Kruger2011} to explore the AMR effects of all textures considered. The energy density contains the exchange energy, the anisotropy energy, the applied field (Zeeman) energy, the magnetostatic (demagnetization) energy and the DMI energy terms. In all simulations, the thickness of the magnetic geometry was $1\,n\mathrm{m}$. Magnetic parameters used in the simulations: saturation magnetization $M_S=956\,k\mathrm{A/m}$, exchange stiffness $A_{ex}=10\,p\mathrm{J/m}$, bulk DMI constant $D=3\,m\mathrm{J}/\mathrm{m}^2$, perpendicular magnetic anisotropy constant $K_u=0.717\,M\mathrm{J}/\mathrm{m}^3$, and applied magnetic field $B_{\mathrm{ext}}=400\,m\mathrm{T}$. The Gilbert damping coefficient $\alpha$ was set to $0.2$. To pin the skyrmions and preclude them from displacing, we introduced magnetic inhomogeneities by tessellating the entire geometry with Gaussian distributed grains with an average size of $10\,n\mathrm{m}$ and allowed for fluctuations in the DMI constant with a $40\%$ variance. For ultrathin layers of a couple atoms in thickness, local thickness variations of as little as one atom can dramatically change thickness-dependent anisotropies, DMI, and so forth, by up to $75\%$~\cite{Bacani2019}. This approach has been used previously to model creep motion of skyrmions in Pt/Co/Ir multilayer stacks~\cite{Legrand2017}.

The magnetization dynamics are obtained numerically by solving the LLG equation for the unit rescaled magnetization $\mathbf{m}= \mathbf M/M_s$  
with spin-transfer-torque effects:\cite{Berger1996,Slonczewski1996}

\begin{align}
(\partial_{t} + \xi\, \mathbf j[U,\mathbf m] \cdot \mathbf{\nabla} ) \mathbf{m} =& -\gamma \mathbf{m} \times \mathbf({B}_{\mathrm{eff}}+{B}_{\mathrm{Th}})\\ \notag
& +\alpha \mathbf{m} \times (\partial_{t} + \frac{\beta}{\alpha} \xi\, \mathbf j[U,\mathbf m] \cdot \mathbf{\nabla}) \mathbf{m},
\end{align}
where $\xi= P\mu_B / (eM_S)$, and $M_S$, $P$, $\mu_B$, $e$ are the saturation magnetization, current polarization (taken to be 0.5), Bohr magneton and electron charge respectively. Ambient temperature effects can be included through the ${B}_{\mathrm{Th}}$ thermal field term:

\begin{eqnarray}
\label{eqn:thielenoise}
\langle\mathbf{B}_{\mathrm{Th}}\rangle &=& 0\\
\langle B_{\mathrm{Th},i}(t) B_{\mathrm{Th},j}(t')\rangle &=& 2\alpha D\frac{k_BT}{\gamma M_S}\delta_{i,j}\delta(t-t'),\label{eqn:diffconst}
\end{eqnarray}
where $k_BT$ is the thermal energy and $\langle\cdot\rangle$ represents averaging over noise realizations. The effective field $\mathbf{B}_{\mathrm{eff}}$ is given by $\mathbf{B}_{\mathrm{eff} }= -M_{S}^{-1} (\delta F[\mathbf{m}]/\delta \mathbf{m})$, where the micromagnetic free energy comprising exchange, anisotropy and dipolar interactions is:

\begin{align}
F=& \int \left( A_{\mathrm{ex}} (\nabla \mathbf{m})^{2} + K_{u} (1-m_z^2)
- \frac{\mu_0}{2} M_S \mathbf m \cdot \mathbf H_d(\mathbf{m}) \right) dV \\ \notag
&+\int  D_B \mathbf{m}\cdot (\nabla \times \mathbf{m})\, dV, 
\end{align}
where the last term describes Bloch DMI.\cite{Dzyaloshinsky1958, Moriya1960a, Thiaville2012} For definiteness, the results presented focused on Bloch skyrmion fabrics.

All models are discretized into cubic cells with the constant cell size of $1\times 1\times 1\,n\mathrm{m}^3$ in the simulations. The linear size of the cells is smaller than both the fundamental length scale $A_{ex}/D\simeq 3.3\,n\mathrm{m}$ and the domain wall width $\sqrt{A_{ex}/K_u}\simeq 3.7\; n\mathrm{m}$. With these parameters, we observed the stabilization of skyrmions with an average diameter of $\sim 30\, n\mathrm{m}$ across temperatures ranging from zero to room temperature.

We generally considered a rectangular geometry of planar dimensions $250\times L\,n\mathrm{m}^2$ where $L$ was allowed to vary from $100 n\mathrm{m}-500\,n\mathrm{m}$ (main results in paper are shown for the case $L=500\,n\mathrm{m}$). Electrical nano-contacts were modeled as gold cylinders with a diameter of $1\,n\mathrm{m}$ symmetrically located along the $x$-axis of the geometry at positions $0.1\,L$ and $0.9\,L$. To self-consistently compute the AMR-mediated current density distribution through the magnetic texture we chose to model a system where we define the conductivity $\sigma_0= (1/\rho_{\parallel} + 2/\rho_{\perp})/3$, ratio of non-adiabatic to adiabatic spin-transfer torque $\beta/\alpha=0.02$, and consider an AMR ratio~\cite{Kruger2011, Prychynenko2017} of $a=\frac{2(\rho_{\parallel}-\rho_{\perp})}{\rho_{\parallel} + \rho_{\perp}}=1.0$. The voltage patterns injected through the contacts to excite the texture were varied in frequency from $0.7-5\,G\mathrm{Hz}$ while the voltage amplitude was set to $V=110\tfrac{L}{L_0}\,m\mathrm{V}$ (where $L_0=500\,n\mathrm{m}$), such that simulation results of geometries with different lateral dimensions could be compared. 

For both the time and space-multiplexed pattern recognition, the sampled state data was run through a support vector machine (SVM) classifier with linear kernel for training and testing. The time-multiplexed data consisted of regularly spaced samples of the time-resolved resistance data. For the space-multiplexed data, a $5\times 6$ grid of $20\,n\mathrm{m}^2$ regions with $20\,n\mathrm{m}$ inter-neighbor spacing was used as a mask to calculate the average out-of-plane magnetization components. As only 8 possible three-pulse combinations existed, our training set consisted of 8 different three-pulse sequences. When varying the number $N$ of read-out elements in the space-multiplex tecnique, the $N$ specific elements used where chosen such as to give a homogeneous sampling of the read-out grid.

\section*{Acknowledgments}

This work was funded by the German Research Foundation (DFG) under the Project No.~EV 196/2-1, and the Transregional Collaborative Research Center (SFB/TRR 173) Spin+X. D.P. would like to thank K.~Litzius, J.-V.~Kim, S.~Kreiss and J.~Sun for fruitful discussions.

\section*{Contributions}
K.E.-S., G.~B., and D.~P.\ developed the concepts underlying the project. D.~P.\ performed the numerical tasks of the project. All authors contributed to the writing and editing of the manuscript.

\appendix
\section{Reservoir Training.}
\label{app:A}

Denote by $\mathbf{m}(t)$ the state of the reservoir at time $t$. Training the output layer of a reservoir requires finding the optimal $D\times N$ weight matrix $\mathbf{\hat{W}}$ mapping some sampling $\mathbf{x}\in \mathbb{R}^D$ of $\mathbf{m}$ into a set of $N$ output nodes $\mathbf{y}$, such that the computed output $\mathbf{y}=\mathbf{\hat{W}}\cdot \mathbf{x}$ minimizes the error $E(\mathbf{y},\mathbf{y}_T)$ with respect to some target output $\mathbf{y_T}(t)$:

\begin{equation}
E(\mathbf{y}(t),\mathbf{y}_T(t))=\sqrt{\langle\lvert\lvert\mathbf{y}(t)-\mathbf{y}_T(t)\rvert\rvert^2\rangle},
\end{equation}
where $\lvert\lvert\cdot\rvert\rvert$ stands for the Euclidean norm and $\langle\cdot\rangle$ is an average over all training data.

Each element $y_i=\hat{W}_{i,j}x_j$ of the output vector corresponds to the value of a given output node. The categorization of the output state is typically defined by the sign of each of the outputs $y_i$. As such, choosing $N$ output nodes, one can in principle codify $2^N$ distinct categories. The notion of hyper-planes discussed in the text and prior literature is born out of the mathematical importance that the threshold $y_i=0$ carries. Once the weights in $\mathbf{\hat{W}}$ have been set, the equation $\hat{W}_{i,j}x_j=0$ effectively describes a $D$-dimensional hyperplane in the state space of the reservoir. Since $N$ of these are being specified (consistently with the number of output nodes chosen), \textit{training} involves tesselating the reservoir state space with $N$, $D$-dimensional, hyper-planes whereas \textit{inference} is the task of identifying which tesselation the reservoir state $\mathbf{m}$ is in.

To parallel between this notation and the skyrmion reservoir system presented in this work, we note: 

\begin{itemize}
	\item $\mathbf{m}(t)$ is the evolution of the infinite dimensional magnetic texture over time.
	\item $\mathbf{x}$ is the input vector. For time-multiplexing, it consists consisting of the resistance time traces sampled at precise temporal intervals $\tau$: $R[\mathbf{m}](t+n\tau)$. For space-multiplexing it corresponds to the coarse-grained snapshots of $\mathbf{m}$.
	\item $\mathbf{y}(t)$ is the set of nodes necessary to identify all the categories needed for input data classification. In the case of sine-square $n$-pulse sequences observed, defining $2^n$ possible outputs, $\mathbf{y}$ is a $n$-dimensional vector. 
	\end{itemize}

The two approaches presented are not mutually exclusive and can in fact be used in tandem to massively increase the dimensionality of the output sampling space. This is particularly important for physical reservoirs such as those presented in this work due to the infinite dimensional nature of the reservoir state $\mathbf{m}$ allowing for arbitrarily large sampling spaces to be considered.

\section{Memory and AMR response behavior at varying frequencies.}
\label{app:B}

In this Appendix we provide further AMR response results with the aim to show how they behave as the driving signal frequency is varied around the system's natural dynamical timescale. 

\begin{figure*}
\includegraphics[width=0.9\textwidth]{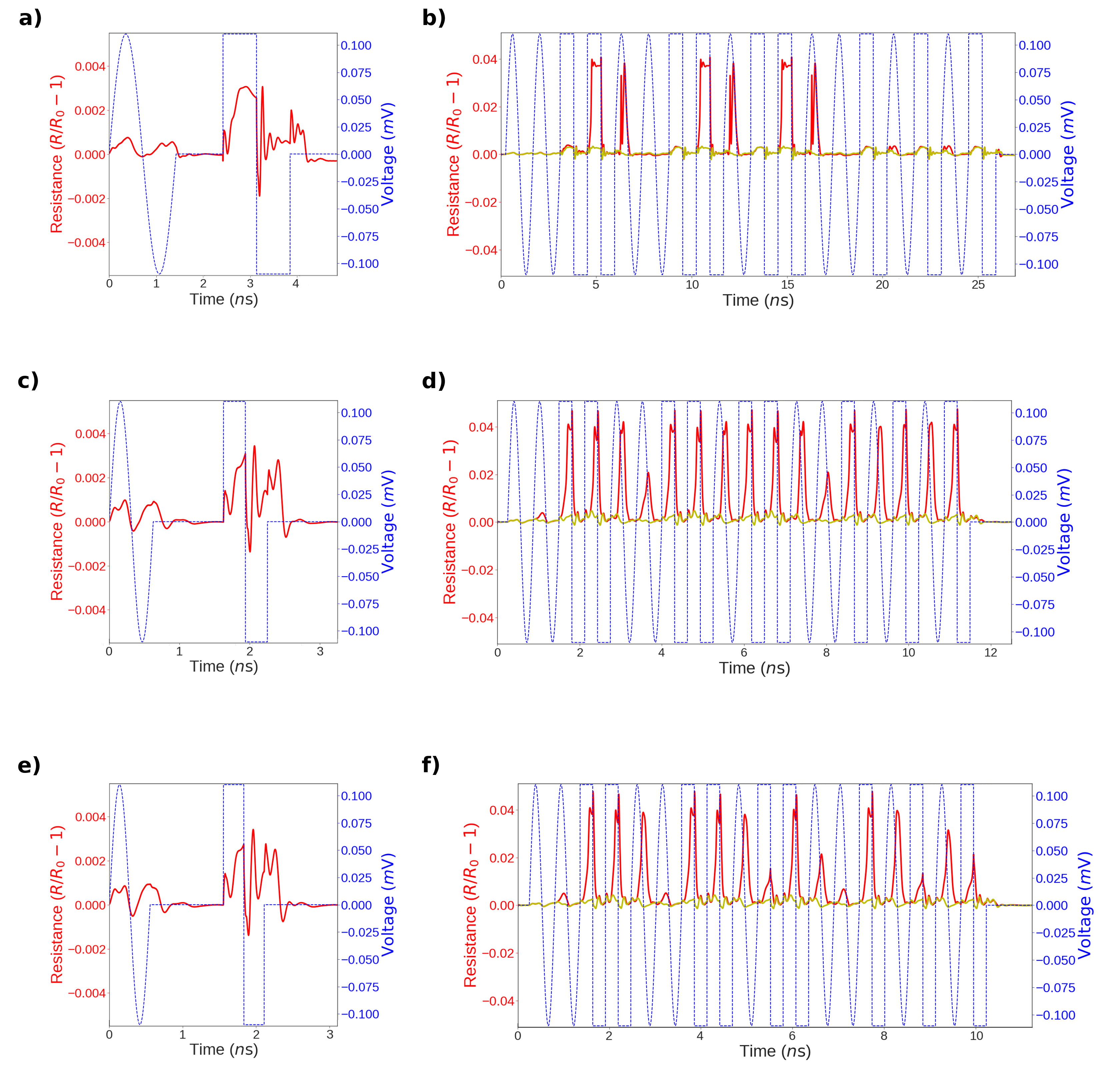}
\caption{Single and pulse train AMR response simulations for varying frequencies as discussed in Section 3: (a,b) $0.7\,G\mathrm{Hz}$, (a,b) $1.6\,G\mathrm{Hz}$, (a,b) $1.8\,G\mathrm{Hz}$. At frequencies far below the system's natural timescale, the magnetization relaxes adiabatically to the driving voltage thus losing all memory of past pulse information. On the other hand, as the frequency is increased above the natural frequency the texture progressively gains more memory of past states. This comes with a tradeoff however as the system eventually ceases to be able to process the driving signals once frequencies get too large.}
\label{fig:7_Frequencies}
\end{figure*}

Figure~\ref{fig:7_Frequencies} plots single and pulse train results (analogous to those discussed in Fig.~\ref{fig:PulseTrain}) for frequencies (a,b) below the system's natural frequency and at frequencies (c,d,e,f) extending above the frequency discussed in the main text ($1\,G\mathrm{Hz}$). At a driving frequency of $0.7\,G\mathrm{Hz}$, the driving signal does not change quickly enough to keep the magnetic texture in a transient dynamical state. As such the texture relaxes adiabatically not retaining much memory of the driving signal's pulse sequence. Brief AMR spikes are seen in accordance with the square pulses. We believe that these might be exaggerated due to the instantaneous rise-time used in the simulations. 

Driving frequencies of $1.6 \,G\mathrm{Hz}$ and $1.8\,G\mathrm{Hz}$ enter a regime which tests the magnetic texture's ability to properly track the driving voltage signal. The pulse train AMR simulations show behavior consistent with the $1\,G\mathrm{Hz}$ simulations discussed in the text. The peak AMR resistance does not change significantly with these slightly larger frequencies further suggesting that the total AMR percentage is set by the sample geometry's dimensions. What is interesting to note in these higher frequency simulations is that the magnetic texture's memory can now respond to more than just the information of two successive pulses. Differently from the $1\,G\mathrm{Hz}$ results shown in Fig.~\ref{fig:PulseTrain})b, the AMR response spikes present more than four distinct shapes. These can be seen to correlate with the state of voltage driving up to three wavelengths into the past. 

As discussed in the main text, at larger frequencies still, the texture decouples from the long-temporal details of driving signal as it begins to respond linearly to the instantaneous pulse driving.

\end{document}